\documentstyle[aps,twocolumn]{revtex}
\begin{document}
\newcommand{\FF}{{\cal F}}
\newcommand{\LL}{{\cal L}}
\newcommand{\OO}{{\cal O}}
\newcommand{\rvac}{|0\rangle}
\newcommand{\lvac}{\langle 0|}
\renewcommand{\=}{&=&} 
\newcommand{\ie}{{{\em i.e.},\ }}
\newcommand{\eg}{{{\em e.g.},\ }}
\renewcommand{\bar}{\overline}
\renewcommand{\a}{\alpha}
\renewcommand{\b}{\beta}
\renewcommand{\d}{\delta}
\newcommand{\g}{\gamma}
\renewcommand{\l}{\lambda}
\newcommand{\m}{\mu}
\newcommand{\n}{\nu}
\renewcommand{\o}{\omega}
\newcommand{\s}{\sigma}
\renewcommand{\t}{\theta}
\newcommand{\tb}{{\bar\theta}}
\newcommand{\aad}{{\a\ad}}
\newcommand{\ad}{{\dot{\alpha}}}
\newcommand{\bd}{{\dot{\beta}}}
\newcommand{\pa}{\partial}
\newcommand{\th}{\widehat\theta}
\newcommand{\yh}{\widehat{y}}
\newcommand{\tbh}{\widehat{\bar\theta}}
\newcommand{\tht}{{\widehat\theta}(t)}
\newcommand{\tbht}{\widehat{\bar\theta}(t)}
\newcommand{\gd}{{\dot{\gamma}}}
\newcommand{\qh}{{\widehat Q}}
\newcommand{\qbh}{{\widehat{\bar Q}}}
\newcommand{\dh}{{\widehat D}}
\newcommand{\dbh}{{\widehat{\bar D}}}
\newcommand{\qb}{{\bar Q}}
\newcommand{\bb}{{\beta}}
\newcommand{\Tr}{{\rm Tr \,}}
\newcommand{\xx}{{\bf x}}
\newcommand{\yy}{{\bf y}}
\newcommand{\id}{{\mathchoice {\rm 1\mskip-4mu l} {\rm 1\mskip-4mu l}
{\rm 1\mskip-4.5mu l} {\rm 1\mskip-5mu l}}}
\newcommand{\real}{{{\rm I} \kern -.19em {\rm R}}}
\newcommand{\y}{(y)}
\newcommand{\pad}[2]{{\displaystyle{\frac{\partial #1}{\partial #2}}}}
\newcommand{\mn}{{\mu\nu}}
\newcommand{\ul}{\underline}
\newcommand{\intq}{{\int d^4 \! x \,\,}}
\newcommand{\half}{\frac 1 2}
\newcommand{\intt}{{\int d^3 \! {\bf x} \,\,}}
\renewcommand{\ss}{\sigma\kern-.54em \sigma}
\newcommand{\be}{\begin{equation}}
\newcommand{\ee}{\end{equation}}
\newcommand{\beq}{\begin{eqnarray}}
\newcommand{\eeq}{\end{eqnarray}}
\newcommand{\ba}{\begin{array}}
\newcommand{\ea}{\end{array}}
\newcommand{\equ}[1]{(\ref{#1})}
\newcommand{\nn}{\nonumber}

\newcommand{\note}[1]{$\bullet\kern-.3em\bullet$[{\bf{#1}}]}
\newcommand{\eqnote}[1]{\bullet\kern-.3em\bullet[{\bf{#1}}]}
\newcommand{\mynote}[1]{$\bullet$[{\bf{#1}}]}
\newcommand{\myeqnote}[1]{\bullet[{\bf{#1}}]}

\title{{\small\begin{flushright}
NEIP--98--009 \\ hep--ph/9808435 \\ August 1998
\end{flushright}}THERMAL SUPERSYMMETRY IN THERMAL SUPERSPACE\thanks{Talk
given at TFT98, The 5th International Workshop on Thermal Field Theories
and their Applications, Regensburg (Germany), August 10-14, 1998.}}
\author{Claudio
Lucchesi\thanks{Supported by the Swiss National Science
Foundation.}\thanks{Email:
claudio.lucchesi@iph.unine.ch}}
\address{Institut de Physique,
Universit\'e de Neuch\^atel, CH--2000 Neuch\^atel, Switzerland}
\maketitle
\begin{abstract}
Thermal superspace is characterized by Grassmann variables which are
time-dependent and antiperiodic in imaginary time, with a period given
by the inverse temperature. The thermal superspace approach allows to
define thermal superfields obeying consistent boundary conditions
and to formulate a ``super-KMS'' condition for superfield propagators.
Upon constructing thermal covariantizations of the superspace derivative
operators, we define thermal covariant derivatives and provide a definition
of thermal chiral and antichiral superfields. Thermal covariantizations
of the generators of the super-Poincar\'e algebra are also constructed,
and the thermal supersymmetry algebra is computed; it has
the same structure as at $T=0$. We then investigate realizations
of this thermal supersymmetry algebra on systems of thermal fields.
In doing so, we observe thermal supersymmetry breaking in terms of the
lifting of the mass degeneracy, and of the non-invariance of the
thermal action.
\end{abstract}

\section{Introduction }\label{cite}

These proceedings present an account of recent work \cite{dl}, in which
realizations of supersymmetry at finite temperature have been
investigated in terms of thermal superfields.
These are defined in a thermally constrained superspace,
baptized ``thermal superspace''.
The thermal superspace approach provides a new framework for the study
of thermal supersymmetry breaking.
Previous investigations of supersymmetry at finite temperature
can be found in \cite{dk}--\cite{bo}.

Supersymmetry and thermal effects are incompatible {\it as is}.
On the one hand, supersymmetry treats bosons and fermions
on an equal footing, as members of the same supermultiplet.
On the other hand, thermal bosons and fermions
are strongly distinguished by their thermal behaviour.
The thermal superspace approach allows to reconcile these
conflicting notions. Thermal superspace is spanned by
{\it time-dependent and antiperiodic Grassmann parameters},
and makes it possible to write consistent boundary conditions, as well as
KMS conditions, at the level of thermal superfields.
These conditions not only can be proven directly
in thermal superspace, they also imply the correct, bosonic or fermionic,
b.c.'s, resp. KMS conditions, for the superfield's components.
So thermal superspace is the correct superspace to be used at finite
temperature, and the information on thermal supersymmetry breaking is
encoded in the temperature-dependent constraints its Grassmann variables
obey. Upon viewing the passage from $T=0$ superspace
to thermal superspace as a change of coordinates, we then
easily construct the thermal covariant derivatives and show that
they provide a consistent definition of the thermal superfields.

At $T=0$, superspace provides a natural representation for the supersymmetry
algebra. Taking the point of view that the same holds at $T>0$, we construct
thermal covariantizations of the supersymmetry generators and compute their
algebra. The thermal supersymmetry algebra so obtained  has the same form
and number of supersymmetries as at $T=0$.
It is only when trying to realize this thermal algebra on thermal
fields that one is faced with thermal supersymmetry breaking.
Indeed, the boundary conditions as well as the KMS relations --
which distinguish bosons from fermions at finite temperature --
are of  space-time {\it global} character.
The supersymmetry algebra, being a local structure, is insensitive
to such global conditions.
At the level of thermal fields however, supersymmetry breaking
is signalled by the lifting of the $T=0$ mass degeneracy, as well as
in terms of the non-invariance of the thermal action under thermal
supersymmetry transformations.

Thermal superspace can be motivated through the
following, heuristic argument. Consider first supersymmetry at zero
temperature. Due to $\{Q_\a,\bar Q_\ad\}=2\s^\m_{\a\ad}P_\m$, the
supersymmetry charges can be viewed as ``square roots'' of translations.
Expressing the supercharges as generalized translations acting through
derivatives only is however not possible in space-time  $x^\m$, the
parameter space of translations, and requires that one enlarges that
parameter space to contain, in addition to $x^\m$, a new set of Grassmannian
coordinates -- denoted $\t$ and $\tb$ -- which are translated under the
action of the supercharges.
A point $X$ in superspace has therefore coordinates
$X = (x^\m,\t^\a,\tb^\ad)$, and since at zero temperature
the parameters of supersymmetry transformations
are space-time constant, the zero-temperature superspace
coordinates  $\t^\a$ and $\tb^\ad$  are space-time constants
as well.

However, one cannot make use of constant  parameters in supersymmetry
transformations rules at finite temperature : the supersymmetry
parameters must be {\it time-dependent} and {\it antiperiodic}
in imaginary time on the interval $[0,\b]$, where $\b =1/T$
denotes the inverse temperature \cite{ggs} (see also
\cite{boyan,fujikawa}).
Adapting straightforwardly the zero-temperature argument above,
it appears natural to require that the variables which are
translated by the effect of the thermal supercharges
bear the same characteristics as the thermal supersymmetry parameters.
{}From this we conclude that thermal superspace must be spanned,
in addition to usual space-time, by Grassmann parameters which are
{\it time-dependent} and {\it antiperiodic}  in imaginary time on
the interval $[0,\b]$.
A point in thermal superspace has therefore coordinates
$$
\widehat X = \Bigl(x^\m,{\widehat\t}^\a(t),{\widehat\tb}^\ad(t)\Bigr)\,,
$$
where a ``hat" is used to denote thermal quantities,
and ${\widehat\t}^\a(t)$, ${\widehat\tb}^\ad(t)$ are subject to
the {\it antiperiodicity conditions}
\be
{\widehat\t}^\a(t+i\b)=-{\widehat\t}^\a(t)\, ,\qquad
{\widehat\tb}^\ad(t+i\b)=-{\widehat\tb}^\ad(t)\,.
\label{antiper}
\ee
These conditions induce a {\it temperature-dependent constraint} on the
{\it time-dependent superspace Grassmann coordinates $\widehat\t(t)$
and $\widehat\tb(t)$}.

The heuristic argument presented here is supported by an
independent formal argument based on the KMS conditions, which
we develop in Section \ref{tfbf}.

\section{From KMS to super-KMS conditions}\label{tfbf}

Consider first a free real scalar field $\varphi$ carrying no conserved
charges. The hamiltonian $H$ being the time evolution operator, the
field $\varphi$ at $x=(t,\xx)$ (in the Heisenberg picture and
with $\hbar=c=1$) can be obtained as
\begin{equation}\label{foh}
\varphi(x)= \varphi(t,{\xx})=e^{iHt}\varphi(0,{\xx})e^{-iHt}\, ,
\end{equation}
with a time coordinate $x^0 =t$ which is allowed to be complex.
The thermal bosonic propagator $D_{C}$ writes
\beq
D_{C}(x_1,x_2)
\= \langle {\sf T}_C \varphi(x_1)\varphi(x_2)\rangle_\bb\nn\\[1mm]
\=\t_C(t_1-t_2)D^>_{C}(x_1,x_2)\nn\\
&&+\t_C(t_2-t_1)D^<_{C}(x_1,x_2)\, ;\nn
\eeq
$\langle ... \rangle_\bb$ denotes the (canonical) thermal average,
${\sf T}_{\! C}$ denotes path ordering, $\t_C$ is the path Heaviside
 function (see \cite{dl} for details),
and the two-point functions $D^>_{C}$, $D^<_{C}$ are defined as
\begin{eqnarray}
D^>_{C}(x_1,x_2)\=\langle \varphi(x_1)\varphi(x_2)\rangle_\bb\, ,\nn
\\[1mm]
D^<_{C}(x_1,x_2)\=\langle\varphi(x_2)\varphi(x_1)\rangle_\bb\, .\nn
\end{eqnarray}
The Boltzmann weight $e^{-\b H}$ can be interpreted as an evolution
operator in imaginary time. Indeed, rewriting \equ{foh} for a
translation in imaginary time by $t=i\beta$, one gets
\be
e^{-\b H} \varphi(t,\xx) e^{\b H}= \varphi(t+i\b,\xx)\,.
\label{bit}
\ee
Now, starting from $D^>_{C}$,
using the cyclicity of the thermal trace (upon inserting $e^{\beta H}
e^{-\beta H} =1$) and the evolution \equ{bit}, one
deduces the {\it bosonic KMS (Kubo-Martin-Schwinger) condition} \cite{kms}.
This condition relates $D_{C}^>$ and $D_{C}^<$ through a translation in
imaginary time:
\begin{equation}\label{kmsb}
D_{C}^> (t_1;\xx_1,t_2;\xx_2) = D_{C}^< (t_1+i\b;\xx_1,t_2;\xx_2)\, .
\end{equation}

A similar derivation holds for fermionic fields.
Defining the fermionic two-point functions  $S^>_{C\, ab}$,
$S^<_{C\, ab}$ [with $a,b=1,...,4$ for Dirac (four-component) spinors]
as
\begin{eqnarray}
S_{C\, {ab}}^{>}(x_1,x_2)
\=\phantom{-}\langle
\psi_a(x_1)\bar\psi_b(x_2)\rangle_\bb\,,\nn\\[1mm]
S _{C\, {ab}}^{<} (x_1,x_2)
\=-\langle\bar\psi_b(x_2)\psi_a(x_1)\rangle_\bb\,,\nn
\end{eqnarray}
and following the same procedure as in the bosonic case, one derives the
{\it fermionic KMS condition}
\begin{equation}\label{kmsf}
S_{C\, {ab}}^{>} (t_1;\xx_1,t_2;\xx_2) =\ - S_{C\, {ab}}^{<}
(t_1+i\b;\xx_1,t_2;\xx_2)\, ,
\end{equation}
which differs from the bosonic one by a relative sign.

We shall be interested ahead in deriving KMS conditions
for superfields. Superfields are usually formulated using two-component
Weyl spinors $\psi_\alpha$ and $\bar\psi^\ad$, which are related
to Dirac spinors through
$
\psi_a = \left(\begin{array}{c} \psi_\alpha \\ \bar\psi^{\ad}
\end{array}\right)\,.
$
The KMS condition for Dirac spinors \equ{kmsf} thus translates into
a set of KMS conditions for two-component spinors $\psi_\a$, $\bar\psi^\ad$.
Defining the thermal two-point functions $S_{C}^{>}$, $S_{C}^{<}$ for
two-component spinors as,
\eg\footnote{This is the only relation we
shall need for practical purposes. Of course, similar relations can
be derived for $S_{C\, \a\b}$, $S_{C\ \ \b}^{\ \, \ad}$
and $S_{C\,}^{\ \, \ad\bd}$.}
\beq
S_{C\,\a}^{>\ \ \dot\b}(x_1,x_2)\=\phantom{-}
\langle\psi_\a(x_1)\bar\psi^\bd(x_2)\rangle_\bb\,,\nn\\
S_{C\,\a}^{<\ \ \dot\b}(x_1,x_2)
\=-\langle\bar\psi^\bd(x_2)\psi_\a(x_1)\rangle_\bb\,,\nn
\eeq
we derive from \equ{kmsf} the {\it fermionic KMS condition for
two-component Majorana spinors}:
\be
\hspace{-3mm}S_{C\,\a}^{>\ \ \dot\b}(t_1;\xx_1,t_2;\xx_2)
= -S_{C\, \a}^{<\ \ \dot\b}(t_1+i\b;\xx_1,t_2;\xx_2)\,.
\label{mmm1}
\ee


The KMS conditions derived above provide an essential, mandatory
characterization of thermal effects at the level of thermal Green's
functions, and induce a clear distinction between bosons and fermions.
So, {\it thermal physics is in conflict with supersymmetry},
which treats bosons and fermions  on an equal footing, as members
of closed supermultiplets. One convenient way of describing supermultiplets
is to use the language of superfields.
Superfields are superspace expansions which contain as components
the bosonic and fermionic degrees of freedom of supermultiplets.
And imposing the KMS conditions for the superfield components at finite
temperature -- if feasible -- must result in temperature-dependent
constraints on the superfield expansion parameters. So we first grant
some freedom to the superspace Grassmann parameters by allowing them
to depend on imaginary time.
Then we show that the superfield formalism can be reconciled with
thermal physics provided superspace is constrained by requiring
that the Grassmann variables be antiperiodic in their imaginary
time variable, with a period given by the inverse temperature.

Let us first settle some notions in the zero-temperature case.
$T=0$ chiral superfields, noted $\phi$, and  $T=0$ antichiral superfields,
denoted by $\bar\phi$, are defined by the conditions
\be
\bar D_\ad \phi =0\,,\qquad\qquad D_\a{\bar\phi} =0\,,
\label{defcs}
\ee
where the $T=0$ covariant derivatives $D_\a$, $\bar D_\ad$ write
\beq
D_\a\= {\partial\over\partial \t^\a}
- i\,\s^\m_{\ \aad}\tb^\ad\,\pa_\m\,,
\label{covder1}\\[1mm]
\bar D_\ad\= {\partial\over\partial \tb^\ad}
-i\,\t^\a\s^\m_{\ \aad}\,\pa_\m\,.
\label{covder2}
\eeq
For chiral and antichiral superfields, it is technically more
convenient to use the so-called chiral and antichiral coordinates,
instead of the usual superspace coordinates $(x^\m,\t,\tb)$.
Chiral, resp. antichiral, coordinates are given by
$(y^\mu,\t^\a,\tb^\ad)$ and  $(\bar y^\mu,\t^\a,\tb^\ad)$, with
$y$ and $\bar y$ defined by the combinations
\beq\label{y}
y^\mu = x^\mu - i\t^\a\sigma^\mu_{\a\ad}\tb^\ad\,, \quad
\bar y^\mu = x^\mu + i\t^\a\sigma^\mu_{\a\ad}\tb^\ad\,.
\eeq
In these variables, the expansions of $T=0$ chiral and antichiral
superfields are simply
\beq
\phi(y,\t)
\= z\y+\sqrt{2}\,\t\,\psi\y-\t\t f\y\, ,\label{chsexp}\\[1mm]
   {\bar\phi}({\bar y},\tb)
\={\bar z}({\bar y})+\sqrt{2}\,\tb\,\bar\psi({\bar y})
   -\tb\tb{\bar f}({\bar y})\, .\label{xexpa}
\eeq
The components of the superfields $\phi$ and $\bar\phi$ form a chiral
supermultiplet which contains two complex scalar fields
$z$ and $f$ and a Majorana spinor
with Weyl components $\psi_\a$ and $\bar\psi^\ad$.

Consider now the $T=0$ chiral-antichiral superfield propagator
$G(y_1,\bar y_2,\t_1,\tb_2)=
\langle 0|T\phi(y_1,\t_1)\bar\phi(\bar y_2,\bar\t_2)|0\rangle$.
Its superspace expansion can equivalently be cast in
the variables $(x,\t,\tb)$ or the chiral ones,
as\footnote{Similar expansions can be written for
the ``chiral-chiral'' superfield propagator
$\langle 0|T\phi(y_1,\theta_1)\phi(y_2,\theta_2)|0\rangle$
and for the conjugate, ``antichiral-antichiral''
 $\langle 0|T\bar\phi(\bar y_1,\bar \t_1)
\bar\phi(\bar y_2,\bar \t_2)|0\rangle$.}
\beq
&&G(y_1,\bar y_2,\t_1,\tb_2)\nn\\[1mm]
&=&D(y_1 -{\bar y}_2 )
-2\t_1^\a\tb_{2\,\bd}\,S_{\a}^{\ \bd} (y_1 -{\bar y}_2 )\nn\\
&&
+\t_1\t_1\,\tb_2\tb_2\,\FF(y_1 -{\bar y}_2 ) \label{cac}\\[1mm]
&=& D(x_1-x_2)
-2\t_1^\a\tb_{2\,\bd}\,S_{\a}^{\ \bd} (x_1-x_2)\nn\\
&&
+\t_1\t_1\,\tb_2\tb_2\,\FF(x_1-x_2)+ \,{\rm derivative\,\,terms}\,,
\nn
\eeq
and includes the $T=0$ Green's functions for the superfield's components:
\beq
D(x_1 -x_2 )&\equiv& \lvac T z(x_1){\bar z}(x_2)\rvac\,,
\nn
\\[1mm]
\FF(x_1 -x_2 )&\equiv&\lvac T f(x_1){\bar f}(x_2 )\rvac\,,
\nn
\\
S_{\a}^{\ \bd} (x_1 -x_2 )&\equiv&\lvac T \psi_\a(x_1)
\bar\psi^\bd(x_2 )\rvac\,.
\nn
\eeq

We now put this system of propagating component fields in a thermal
bath at some finite temperature $T$. As the heat bath  affects
propagation, the $T=0$ propagators above must be replaced
by their thermal counterparts:
\beq
D_C(x_1 -x_2 )
&\equiv&
\langle {\sf T}_{\!C} \  z(x_1){\bar z}(x_2)\rangle_\bb
\nn
\,,
\\[1mm]
\FF_C(x_1 -x_2 )
&\equiv&
\langle {\sf T}_{\!C} \  f(x_1){\bar f}(x_2 )\rangle_\bb
\nn
\,,
\\
S_{C\,\a}^{\ \ \ \bd} (x_1 -x_2 )
&\equiv&
\langle {\sf T}_{\!C} \  \psi_\a(x_1)\bar\psi^\bd(x_2 )\rangle_\bb\,,
\nn
\eeq
which have to  obey the relevant KMS condition.
That is, thermal propagators of scalar components must obey
the bosonic KMS condition \equ{kmsb},
\beq
D_{C}^>(t_1;\xx_1,t_2;\xx_2)
\= \phantom{-}D_{C}^< (t_1+i\b;\xx_1,t_2;\xx_2)
\,,\label{c1}\\[1mm]
\FF_C ^>(t_1;\xx_1,t_2;\xx_2)
\= \phantom{-}\FF_C^< (t_1+i\b;\xx_1,t_2;\xx_2)
\,,\label{c3}
\eeq
while the thermal propagator of the fermionic component
has to satisfy the fermionic constraint \equ{mmm1},
\be
\hspace{-1.5mm}
S_{C\,\a}^{>\ \ \bd} (t_1;\xx_1,t_2;\xx_2)
= -
S_{C\,\a}^{< \ \ \bd} (t_1+i\b;\xx_1,t_2;\xx_2)\,.
\label{c2}
\ee


But the presence of a heat bath not only enforces the KMS conditions for
the superfields components. It also obliges us to {\it adapt the notion
of superfield to the thermal context}. Indeed the usual $T=0$ superfield
formulation is by construction supersymmetric, and is therefore
incompatible with thermal effects, as we have argued above.

In the Introduction, we have motivated the fact that, at finite
temperature, the superspace Grassmann variables should be dependent
on imaginary time and antiperiodic. Consequently, we promote $\t$
and $\tb$ to become thermal superspace coordinates
$\t\rightarrow\th=\th(t)$, $\tb\rightarrow\tbh=\tbh(t)$ obeying
the antiperiodicity properties \equ{antiper}.

In taking superspace Grassmann coordinates
$\th(t)$, $\tbh(t)$, we also introduce a formal time-dependence
in the second terms of the variables $y$ and ${\bar y}$
\equ{y},  which we henceforth denote by:
\be
\widehat y^\mu_{(t)}
=
x^\m -i\widehat\t(t)\s^\m\widehat\tb(t)\,,
\quad
\widehat{\bar y}^\mu_{(t)}
=
x^\m +i\widehat\t(t)\s^\m\widehat\tb(t)\,.
\nn
\ee


In thermal superspace, we define chiral and antichiral superfields at
finite temperature, denoted by the ``hat" notation $\widehat\phi$, resp.
$\widehat{\bar\phi}$, similarly to \equ{chsexp}, \equ{xexpa}, but with
{\it the thermal superspace Grassmann coordinates $\th(t)$, $\tbh(t)$}
as the expansion parameters. This yields\footnote{Here
and in the sequel, we  simplify the notation
by occasionally using $\widehat y_i$ and $\th_i$ instead of
$\widehat y_i(t_i)$ and $\th_i(t_i)$ in non ambiguous
situations.}
\beq
\widehat\phi[\widehat y,\widehat\t]
\=
z[\widehat y]+\sqrt{2}\,\widehat\t\,\psi[\widehat y]
- \widehat\t\widehat\t\, f[\widehat y]\,,
\label{tch}\\[1mm]
\widehat{\bar\phi}\,[\widehat{\bar y},\widehat\tb]
\=
{\bar z}[\widehat{\bar y}]+\sqrt{2}\,\,\widehat\tb\,
\bar\psi[\widehat{\bar y}]
- \widehat\tb\widehat\tb\,
{\bar f}[\widehat{\bar y}]\,.
\label{tach}
\eeq
[The consistency of these thermal expansions is discussed in Section
\ref{tcd} ahead, where we also give the $(x,\th,\tbh)$-expansion for
$\widehat\phi$, eq. \equ{along}].
In the same spirit, we next define the thermal chiral-antichiral
superfield propagator
$
G_C\,[\widehat y_{1}, \widehat{\bar y}_{2},\widehat\t_1,
\widehat\tb_2]
\equiv
\langle {\sf T}_{\!C}\  \widehat\phi[\widehat y_{1},\widehat\t_1]
\,\widehat{\bar\phi}\,[\widehat{\bar y}_{2},\widehat\tb_2] \rangle_\bb$
and expand it in thermal superspace, in analogy to \equ{cac}, as
\beq
&&G_C[\widehat y_{1(t_1)}, \widehat{\bar
y}_{2(t_2)},\widehat\t_1(t_1),\widehat\tb_2(t_2)]\nn\\[1mm]
&=&\!  D_C[\widehat y_{1(t_1)}\! -\! \widehat{\bar y}_{2(t_2)}]
-2\widehat\t_1^\a(t_1)\tb_{2\,\bd}(t_2)S_{C\,
\a}^{\ \ \ \bd}[\widehat y_{1(t_1)}\! -\! \widehat{\bar y}_{2(t_2)}]
\nonumber \\
&&  + \ \widehat\t_1(t_1)\widehat\t_1(t_1)\,
\widehat\tb_2(t_2) \widehat\tb_2(t_2)\,\FF_C [\widehat y_{1(t_1)} -
\widehat{\bar y}_{2(t_2)}]
\nonumber
\eeq
The thermal superfield two-point functions $G_C^>$, resp. $G_C^<$, can be
defined in relation to $G_C$ through
\beq
&&G_{C}\,[\widehat y_{1(t_1)},
\widehat{\bar y}_{2(t_2)},\widehat\t_1(t_1),\widehat\tb_2(t_2)]\nn\\[1mm]
&=& \t_C(t_1-t_2) \,
G^>_{C}\,[\widehat y_{1(t_1)}, \widehat{\bar y}_{2(t_2)},
\widehat\t_1(t_1),\widehat\tb_2(t_2)]\nn\\
&&+\ \t_C(t_2-t_1)\, G^<_{C}\,[\widehat y_{1(t_1)},
\widehat{\bar y}_{2(t_2)},\widehat\t_1(t_1),\widehat\tb_2(t_2)]\,,
\nn
\eeq
with
\beq
G_C^>\,[\widehat y_{1}, \widehat{\bar
y}_{2},\widehat\t_1,\widehat\tb_2]\=\langle
\widehat\phi[\widehat y_{1},\widehat\t_1]\,\widehat{\bar\phi}\,[
\widehat{\bar y}_{2},\widehat\tb_2] \rangle_\bb
\,,\label{asup}\\[1mm]
G_C^<\,[\widehat y_{1}, \widehat{\bar
y}_{2},\widehat\t_1,\widehat\tb_2]\=\langle
\widehat{\bar\phi}
[\widehat{\bar y}_{2},\widehat\tb_2]
\,\widehat\phi[\widehat y_{1}, \widehat\t_1]
\rangle_\bb\label{rsup}\,.
\eeq

The KMS condition can now be formulated at the level of thermal superfield
propagators. The {\it superfield KMS (or super-KMS) condition} is:
\beq
&&G_C^>\,[\widehat y_{1(t_1)}, \widehat{\bar
y}_{2(t_2)},\widehat\t_1(t_1),\widehat\tb_2(t_2)]\nn\\[1mm]
&=&
G_C^<\,[\widehat y_{1(t_1 + i\beta)}, \widehat{\bar
y}_{2(t_2)},\widehat\t_1(t_1+i\b),\widehat\tb_2(t_2)]\,,
\label{skms}
\eeq
with the time-translated variable $\widehat y_{1(t_1 + i\beta)}$ given by
\be
\widehat y_{1(t_1 + i\beta)} =
\widehat y_{1(t_1)} + (i\beta\,\,;\,\,{\bf 0})\, ,
\label{y1}
\ee
upon making use of the antiperiodicity \equ{antiper}.


Clearly, the superfield KMS condition \equ{skms} is of bosonic type, since
chiral and antichiral superfields are bosonic objects.
This condition can be proven at the superfield level in a way similar
to the case of the scalar field.
Let us start by formulating the evolution in imaginary time for
a, \eg chiral,  thermal superfield. Appling this evolution, eq.
\equ{bit}, to the components in $\widehat\phi$, and using \equ{y1},
we get
\beq
&&e^{-\b H} \, \widehat\phi[\widehat y_{(t)},\widehat\t(t)] \, e^{\b H}
\nn\\[1mm]
\=
z[\widehat y^0_{(t)}+i\beta ; \widehat{\vec y}_{(t)}]
+\sqrt{2}\,\widehat\t(t)\,\psi[\widehat y^0_{(t)}+i\beta ;
\widehat{\vec y}_{(t)}]
\nn\\
&&-\widehat\t(t)\widehat\t(t)\,f[\widehat y^0_{(t)}+i\beta ;
\widehat{\vec y}_{(t)}]
\nn\\[1mm]
\=
\widehat\phi[\widehat y_{(t_1 + i\beta)},\widehat\t(t)]
\,.
\label{sbit22}
\eeq
Note that the time arguments of $\widehat\t(t)$ and $\widehat\tb(t)$
are {\it not} shifted. The thermal Grassmann variables -- which are
{\it coordinates} -- do not undergo dynamical evolution in imaginary time
under the hamiltonian.

In order to prove the superfield KMS relation \equ{skms},
we start from the thermal superfield two-point function $G_C^>$
\equ{asup},
\beq
&&G_C^>\,[\widehat y_{1(t_1)}, \widehat{\bar
y}_{2(t_2)},\widehat\t_1(t_1),\widehat\tb_2(t_2)]\nn\\[1mm]
\=
{1\over Z(\b)}\, \Tr \Biggl\{ e^{- \beta H}
\widehat\phi
[\widehat y_{1(t_1)},\widehat\t_1(t_1)]\,
\widehat{\bar\phi}\,
[\widehat{\bar y}_{2(t_2)},\widehat\tb_2(t_2)] \Biggr\}\,,\nn
\eeq
and introduce the thermal component expansions for the superfields [eqs.
\equ{tch}--\equ{tach}]. We then rotate cyclically $\widehat{\bar\phi}$ to
the front, insert the identity $e^{\beta H}e^{-\beta H}=1$, and  rotate
$e^{-\b H}$ to the front. The right side therefore rewrites as:
\beq
&& \displaystyle{1\over Z(\b)} \Tr\! \Biggl\{e^{-\beta H}
\Bigl({\bar z}[\widehat {\bar y}_2]-\sqrt{2}\widehat \tb_2
\bar\psi[\widehat {\bar y}_2]-\widehat \tb_2\widehat\tb_2\,
{\bar f}[\widehat {\bar y}_2]\Bigr)
\nn\\
&&\hspace{3mm}\times
 e^{- \beta H} \Bigl(z[\widehat y_1]+\sqrt{2}\widehat \t_1\psi[\widehat
y_1]-\widehat \t_1
\widehat\t_1\, f[\widehat y_1]\Bigr) e^{\beta H} \Biggr\}.\nn
\eeq
The negative sign in front of the fermionic component of
 $\widehat{\bar\phi}$ follows from
the anticommutativity of the Grassmann variables.  We now insert the
superfield time evolution \equ{sbit22} and recast  the  last expression as:
\beq
&&\displaystyle{1\over Z(\b)}\,
\Tr
\Biggl\{
e^{- \beta H}
\Bigl({\bar z}[\widehat{\bar y}_{2(t_2)}]
-\sqrt{2}\,\,\widehat\tb_2(t_2)\,\bar\psi[\widehat{\bar y}_{2(t_2)}]
\nn\\[1mm]
&&-\widehat\tb_2(t_2)\widehat\tb_2(t_2)\,
        {\bar f}[\widehat{\bar y}_{2(t_2)}]\Bigr)
\times
\Bigl(z[\widehat y_{1(t_1+i\b)}]  \nn\\
&&+\sqrt{2}\,\widehat\t_1(t_1)\,\psi[ \widehat y_{1(t_1+i\b)}]
-\widehat\t_1(t_1)\widehat\t_1(t_1)\,
        f[ \widehat y_{1(t_1+i\b)}]\Bigr) \Biggr\}\,.\nn
\eeq
Making  use of the antiperiodicity \equ{antiper}
of the Grassmann variables, we set
$\widehat\t_1(t_1)=- \widehat \t_1(t_1 +i\beta)$ and get
\beq
&&\displaystyle{1\over Z(\b)}\, \Tr \Biggl\{ e^{- \beta H}
\Bigl({\bar z}[\widehat{\bar y}_{2(t_2)}]-\sqrt{2}\,\,\widehat\tb_2(t_2)\,
\bar\psi[\widehat{\bar y}_{2(t_2)}]
\nn\\[-1mm]
&&\phantom{\displaystyle{1\over Z(\b)}\, \Tr \Biggl\{}
-\widehat\tb_2(t_2)\widehat\tb_2(t_2)\,
{\bar f}[\widehat{\bar y}_{2(t_2)}]\Bigr)
\nn\\[-1mm]
&&\hspace{3mm}\times
\Bigl(z[\widehat y_{1(t_1+i\b)}]  -
\sqrt{2}\,\widehat\t_1(t_1+i\b)\,\psi[\widehat
y_{1(t_1+i\b)}]
\nn\\[-1mm]
&&\phantom{\displaystyle{1\over Z(\b)}\, \Tr \Biggl\{}
-\widehat\t_1(t_1+i\b)\widehat\t_1(t_1+i\b)\,
f[\widehat y_{1(t_1+i\b)}]\Bigr)\Biggr\}\, .\nn
\eeq
Observing that fermionic fields do not propagate into bosonic fields,
and {\it vice-versa}, we replace the two negative signs in front of the
fermionic components by two positive signs. The second thermal superfield
in the product then identifies to  $\widehat\phi$ \equ{tch} with all time
arguments shifted by $i\b$.
As a result, the above computation yields just the  superfield KMS
condition \equ{skms}, which is hereby proved.


We verify now that the superfield KMS condition \equ{skms} yields
the expected component relations \equ{c1}--\equ{c3} and \equ{c2}.
Expanding $\widehat\phi$ and $\widehat{\bar\phi}$ along
\equ{tch}--\equ{tach} in eqs. \equ{asup}--\equ{rsup} yields
\beq
&&G_C^>\,[\widehat y_{1(t_1)}, \widehat{\bar
y}_{2(t_2)},\widehat\t_1(t_1),\widehat\tb_2(t_2)]\nn\\
\=
D_C^>\,[\widehat y_{1(t_1)}, \widehat{\bar y}_{2(t_2)}]
- 2\widehat\t_{1}^\a(t_1)\widehat\tb_{2\, \bd}(t_2) \,S_{C\, \a}^{>\,\
\bd}\,[\widehat y_{1(t_1)},
\widehat{\bar y}_{2(t_2)}] \nn\\
&&+ \widehat\t_1(t_1)\widehat\t_1(t_1)
\widehat\tb_2(t_2)\widehat\tb_2(t_2)
\,\FF^>_C[\widehat y_{1(t_1)}, \widehat{\bar y}_{2(t_2)}]\,,
\nn
\eeq
and
\beq
&&G_C^<\,[\widehat y_{1(t_1)}, \widehat{\bar
y}_{2(t_2)},\widehat\t_1(t_1),\widehat\tb_2(t_2)]\nn\\
\=
D_C^<\,[\widehat y_{1(t_1)}, \widehat{\bar y}_{2(t_2)}]
- 2\widehat\t_{1}^\a(t_1)\widehat\tb_{2\, \bd}(t_2) \,
S_{C\, \a}^{<\,\ \bd}\,[\widehat y_{1(t_1)},
\widehat{\bar y}_{2(t_2)}] \nn\\
&&+ \widehat\t_1(t_1)\widehat\t_1(t_1)
\widehat\tb_2(t_2)\widehat\tb_2(t_2)\, \FF^<_C[\widehat
y_{1(t_1)},
\widehat{\bar y}_{2(t_2)}]\, .
\nn
\eeq
Repalcing these developments in the superfield KMS condition \equ{skms}
leads then to the following:
{\bf (i)} For the scalar component,
$$
D_C^>\,[\widehat y_{1(t_1)}, \widehat{\bar y}_{2(t_2)}] =
D_C^<\,[\widehat y_{1(t_1 + i\b)}, \widehat{\bar y}_{2(t_2)}],
$$
which reduces, when returning to the variables $x=(t;\xx)$ by Taylor
expanding around $x^\m$, to the bosonic KMS relation \equ{c1}.
{\bf (ii)} For the fermionic component,
\beq
&&\widehat\t_1^\a(t_1)\widehat\tb_{2\, \bd}(t_2) \,S_{C\, \a}^{>\,\
\bd}\,[\widehat y_{1(t_1)},
\widehat{\bar y}_{2(t_2)}]\nn\\
\=
\widehat\t_1^\a(t_1+i\beta)\widehat\tb_{2\, {\bd}}(t_2) \,
S_{C\, \a}^{<\,\ \bd}\,[\widehat y_{1(t_1 + i\b)} ,
\widehat{\bar y}_{2(t_2)}].
\nn
\eeq
With the antiperiodicity condition \equ{antiper},
$\widehat\t_1^\a(t_1+i\beta)=-\widehat\t_1^\a(t_1)$, we obtain
$$
S_{C\, \a}^{>\,\ \bd}\,[\widehat y_{1(t_1)}, \widehat{\bar y}_{2(t_2)}]
=
- S_{C\, \a}^{<\,\ \bd}\,[\widehat y_{1(t_1 + i\b)},
\widehat{\bar y}_{2(t_2)}]\,,
$$
which yields, in the variables $x=(t;\xx)$, the fermionic KMS condition
\equ{c2} with the correct relative sign.
Finally {\bf (iii)} for the auxiliary field, one gets
\beq
&&\widehat\t_1(t_1)\widehat\t_1(t_1)
\widehat\tb_2(t_2)\widehat\tb_2(t_2)
\,\FF^>_C[\widehat y_{1(t_1)},
\widehat{\bar y}_{2(t_2)}]\nn\\
\=
\widehat\t_1(t_1\!+\!i\b)\widehat\t_1(t_1\!+\!i\b)
\widehat\tb_2(t_2)\widehat\tb_2(t_2)\,
\FF^<_C[\widehat y_{1(t_1\! +\! i\b)},
\widehat{\bar y}_{2(t_2)}].
\nn
\eeq
With $\widehat\t_1(t_1 + i\b)= -\widehat\t_1(t_1)$ \equ{antiper} and
in the variables $x=(t;\xx)$, this is the bosonic KMS condition \equ{c3}.

\section{Thermal covariant derivatives}\label{tcd}

Deriving expressions for the covariant derivatives and supercharges  on
thermal superspace can be done simply by performing the change of variables
from usual, zero temperature, superspace to thermal superspace, \ie
$$
(x^\m,\t,\tb)\quad \longrightarrow\quad
\left(x'^\m=x^\m,\t'=\th(t),\tb'=\tbh(t)\right)\, ,
$$
with $t=x^0$. Under this change of variables, the partial derivatives with
respect to $\xx$, $\t$ and $\tb$ transform trivially,
\beq
\pad{}{\xx} &\longrightarrow& \pad{}{\xx'} = \pad{}{\xx}
\, ,\nn\\[1mm]
\pad{}{\t^\a}&\longrightarrow&\pad{}{\t'^\a}=\pad{}{\th^\a}
\,,\nn\\[1mm]
\pad{}{\tb^\ad}&\longrightarrow&\pad{}{\tb'^\ad}=\pad{}{\tbh^\ad}
\,,\nn
\eeq
while the time derivative has to take the time-dependence of the thermal
Grassmann variables into account:
$$
\pad{}{t} \longrightarrow \pad{}{t'} + \pad{\th^\a}{t}\pad{}{\th^\a} +
\pad{\tbh^\ad}{t}\pad{}{\tbh^\ad}\,, \qquad
\left(\pad{t'}{t}=1\right)\, .
$$
Consequently, we define the partial time derivative  at finite temperature
as
\be
\widehat{\pad{}{t}} \ \equiv \ \pad{}{t} - \Delta\,,\quad
\Delta = \pad{\th^\a}{t}\pad{}{\th^\a} + \pad{\tbh^\ad}{t}\pad{}{\tbh^\ad}\,.
\label{finttt}
\ee
We call this object the {\it thermal time derivative};  $\Delta$ accounts
for the thermal corrections. Accordingly, we also define a {\it thermal
space-time derivative} as
\be
\widehat\pa_\m=
\left( \pad{}{t} - \Delta \ \ ; \ \pad{}{\xx}\right)\,.
\label{pamch}
\ee

To construct the thermal covariant derivatives, we replace
 in the expressions of the zero-temperature  covariant derivatives
\equ{covder1}--\equ{covder2}  the $T=0$ Grassmann variables and
derivative operators by their thermal counterparts.
This means that (i) we replace the
zero-temperature, constant Grassmann parameters $\t$, $\tb$ by the thermal,
time-dependent and antiperiodic parameters $\th$, $\tbh$, and that (ii) the
derivative operators $\pa_\m$, $\pa/\pa{\t}$ and $\pa/\pa{\tb}$ are replaced
by $\widehat\pa_\m$, $\pa/\pa{\th}$ and $\pa/\pa{\tbh}$. This yields {\it
thermal covariant derivatives} $\widehat D_\a$ and $\widehat{\bar D}_\ad$
in the form:
\beq
\widehat D_\a \= \pad{}{\th^\a} -i\,\s^\m_\aad\tbh^\ad \pa_\m
+i\,\s^0_\aad\tbh^\ad \Delta\,,
\nn\\[1mm]
\widehat{\bar D}_\ad \= \pad{}{\tbh^\ad} -i\, \th^\a\s^\m_\aad \pa_\m
 +i\, \th^\a\s^0_\aad \Delta\,.\nn
\eeq

In order to validate these expressions, we observe that they play, in
thermal superspace, the same role as the usual covariant derivatives of
supersymmetry in $T=0$ superspace.

First, the thermal covariant derivatives obey the same anticommutation
relations as at $T=0$. This can readily be checked by direct computation of
the anticommutators. One obtains, in perfect analogy to the $T=0$ case,
$$
\{ \widehat D_\alpha , \widehat{\bar D}_\ad \} =
-2i\sigma^\mu_{\alpha\ad}\widehat\partial_\mu\,,\quad
\{\widehat D_\a\,,\, \widehat D_\b \} = \{ \widehat{\bar D}_\ad \,,\,
\widehat{\bar D}_\bd \} =0\,.
$$
This is actually obvious upon noticing that the thermal space-time derivative
$\widehat\pa_\m$ gives zero when acting on the $t$-dependent variables $\th$,
$\tbh$, since
\be
\ba{l}
\widehat\pa_0\,\th^\a= \pad{\th^\a}{t} - \pad{\th^\g}{t}\delta^\a_\g =0
\,,\\[3mm]
\widehat\pa_0\,\tbh^\ad= \pad{\tbh^\ad}{t} -
\pad{\tbh^\gd}{t}\delta^\ad_\gd =0\,.
\ea
\label{obs1}
\ee
Therefore $\widehat\pa_\m$ plays the same role for the thermal
Grassmann variables $\th$, $\tbh$ as that of the usual space-time derivative
$\pa_\m$ for the $t$-independent, non thermal $\t$ and $\tb$.
In this sense, the thermal time (and consequently the thermal space-time)
derivative is a covariantization, with respect to thermal superspace, of
the zero-temperature time (space-time) derivative.

Second, the thermal covariant derivatives $\widehat{\bar D}_\ad$ and
$\widehat D_\a$ provide a {\it definition} of the thermal chiral and
antichiral superfields $\widehat\phi$, $\widehat{\bar\phi}$, eqs. \equ{tch}
and \equ{tach}, as the solution to the thermal generalization of the
conditions \equ{defcs}:
$$
\widehat{\bar D}_\ad\ \widehat\phi = 0\,,\qquad \qquad
\widehat D_\a\ \widehat{\bar\phi}= 0\, .
$$

Our thermal superfield expansions \equ{tch} and \equ{tach} can easily be
seen to be consistent also from the point of view of the fields'
boundary conditions.
Thermal, \eg chiral, superfields being bosonic objects, they must
obey a {\it superfield periodic boundary condition} in the form:
$$
\widehat\phi[\widehat y_{(t)},\th(t)]
=\widehat\phi[\widehat y_{(t+i\b)},\th(t+i\b)]\,.
$$
In the variables $(x,\th,\tbh)$, $x=(t;\xx)$, this condition writes
\be
\widehat\phi[t;\xx,\th(t),\tbh(t)]=
\widehat\phi[t+i\b;\xx,\th(t+i\b),\tbh(t+i\b)]\,.
\label{ssbc}
\ee
Expanding now both sides in thermal superspace along
\beq
\widehat\phi[x,\th,\tbh]\=
z(x) +\sqrt{2}\th\psi(x)-\th\th f(x) -i(\th\sigma^\m\tbh)\pa_\m z(x)\nn\\
&&+{i\over\sqrt{2}}\th\th([\pa_\m\psi(x)]\sigma^\m\tbh)
-{1\over 4} \th\th\tbh\tbh\Box z(x)
\,,
\label{along}
\eeq
we immediately get from \equ{ssbc} {\bf(i)} for the scalar field $z$ the
periodic b.c.
\be
z(t;\xx)=z(t+i\b;\xx)\ ,
\label{pbc1}
\ee
{\bf (ii)} for the fermionic field $\psi$, upon replacing
$\th(t+i\b)=-\th(t)$ [eq. \equ{antiper}], the antiperiodic b.c.
\be
\psi(t;\xx)=-\psi(t+i\b;\xx)\ ,
\label{abc}
\ee
and {\bf (iii)} for the scalar field $f$, with
$\th(t+i\b)\th(t+i\b)=\th(t)\th(t)$ [eq. \equ{antiper}], the periodic b.c.
\be
f(t;\xx)=f(t+i\b;\xx)\ ,
\label{pbc2}
\ee
as well as additional conditions for the fields' derivatives.

\section{Thermal covariantization of the supersymmetry algebra}\label{tssa}

The main interest of superspace lies in
the natural representation it provides for the super-Poincar\'e algebra
in terms of derivatives with respect to superspace coordinates.
The purpose of this section is to construct the supersymmetry
generators acting on thermal superspace, and to compute their algebra.
However, {\it the existence of a supersymmetry algebra on thermal superspace
should not be assimilated to a statement that supersymmetry does not break at
 finite $T$}. That such an algebra exists does  {\it not} imply that a
supersymmetric field theory can be constructed carrying the same symmetry
algebra.

The zero-temperature supercharges are, in our conventions:
\beq
Q_\a \= -i\,\pad{}{\t^\a} + \s^\m_\aad\tb^\ad \pa_\m\, ,
\nn\\[1mm]
\bar Q_\ad \=\phantom{-} i\,\pad{}{\tb^\ad} - \t^\a\s^\m_\aad \pa_\m\,.
\nn
\eeq
The corresponding thermal objects are constructed using the same procedure as
the one used above for the thermal covariant derivatives, that is, we replace
 $\t$, $\tb$ by $\th$, $\tbh$, and $\pa_\m$, $\pa/\pa{\t}$, $\pa/\pa{\tb}$
by $\widehat\pa_\m$ [eq. \equ{pamch}], $\pa/\pa{\th}$ and $\pa/\pa{\tbh}$.
This yields the
following expressions for the {\it thermal supercharges}:
\beq
\qh_\a \= -i\pad{}{\th^\a} + \s^\m_\aad\tbh^\ad \pa_\m \nn\\
&&-\s^0_\aad\tbh^\ad
\left( \pad{\th^\g}{t}\pad{}{\th^\g} +
\pad{\tbh^\gd}{t}\pad{}{\tbh^\gd} \right)
\, ,\label{ths}\\[1mm]
\qbh_\ad \= \phantom{-} i\pad{}{\tbh^\ad} - \th^\a\s^\m_\aad \pa_\m \nn\\
&&+\th^\a\s^0_\aad
\left(
 \pad{\th^\g}{t}\pad{}{\th^\g} +
\pad{\tbh^\gd}{t}\pad{}{\tbh^\gd} \right)\,.\label{thsb}
\eeq
or, in a compact form,
\beq
\qh_\a \= -i\,\pad{}{\th^\a} + \s^\m_\aad\tbh^\ad \widehat\pa_\m
\,,\nn\\[1mm]
\qbh_\ad \= \phantom{-}i\,\pad{}{\tbh^\ad} - \th^\a\s^\m_\aad \widehat\pa_\m
\, .\nn
\eeq
It is straightforward to check that thermal supercharges obey the same
anticommutation relations with thermal covariant derivatives as at $T=0$:
$$
\{ \qh_\a \,,\, \dh_\b \} =
\{ \qbh_\ad \,,\, \dh_\b \} =
\{ \qh_\a \,,\, \dbh_\bd \} =
\{ \qbh_\ad \,,\, \dbh_\bd \} = 0\, .
$$
Furthermore we have
$$
\{ \widehat Q_\alpha , \widehat{\bar Q}_\ad \} =
2i\sigma^\mu_{\alpha\ad}\widehat\partial_\mu,\quad
\{ \qh_\a \,,\, \qh_\b \} = \{ \qbh_\ad \,,\, \qbh_\bd \}=0\, .
$$


In order to compute the full thermal super-Poincar\'e algebra, we need in
addition expressions for the thermal translations and thermal Lorentz
generators.
At finite temperature, the translation and Lorentz generators above are
modified -- similarly to the thermal covariant derivatives and the thermal
supercharges -- by replacing $\t$, $\tb$ by $\th$, $\tbh$, and $\pa_\m$,
$\pa/\pa{\t}$, $\pa/\pa{\tb}$ by $\widehat\pa_\m$ [eq. \equ{pamch}],
$\pa/\pa{\th}$ and $\pa/\pa{\tbh}$.
Therefore we define the action of {\it
thermal translation} and {\it thermal Lorentz generators} on a thermal scalar
superfield through
\beq
[ \widehat P^\m , \widehat \phi(x,\th,\tbh)] \=
-i\,\widehat \pa^\m\widehat \phi(x,\th,\tbh)\,,
\nn\\[1mm]
[\widehat M^\mn,\widehat \phi(x,\th,\tbh)]
 \=
\Biggl[i(x^\m\widehat \pa^\n -x^\n\widehat \pa^\m)\nn\\[-1mm]
+ \displaystyle{i\over 2}
(\s^\mn)_\a^{\ \b}\th_\b&&\!\!\!\!\!\!
\pad{}{\th_\a}
+ \displaystyle{i\over 2}\,
(\bar\s^{\n\m})^\ad_{\ \bd}\tbh^\bd\pad{}{\tbh^\ad}\Biggr]
\widehat\phi(x,\th,\tbh)\,.\nn
\eeq

As only the derivative in the time direction is modified at finite
temperature, we now distinguish between the generators which are genuinely
thermal [that is, which involve the operator $\Delta$ in \equ{finttt}] and
those generators of which the only thermal character comes from the
superspace coordinates being the thermal ones. We drop the ``hat" for the
latter operators, and hence decompose the thermal translations
$\widehat P^\m$ into thermal time translations $\widehat P^0$ and space
translations $P^i$, while the thermal Lorentz generators $\widehat M^{\mn}$
are separated into thermal Lorentz boosts $\widehat M^{0i}$ and space
rotations $M^{ij}$.
A straightforward computation of the commutation rules yields the thermal
Poincar\'e algebra -- the bosonic sector of the thermal super-Poincar\'e
algebra:
\beq
[\widehat M^{0i},\widehat P^0]\=-i\, P^i\, ,
\nn\\[1mm]
[\widehat M^{0i}, P^j]\= \phantom{-} i\, \eta^{ij} \widehat P^0\, ,
\nn\\[1mm]
[M^{ij},\widehat P^0]\=\phantom{-}0\, ,
\nn\\[1mm]
[M^{ij},P^k]\=-i\,(\eta^{ik} P^j - \eta^{jk} P^i)\, ,
\nn\\[1mm]
[M^{ij}, M^{kl}]\= -i\,( \eta^{ik}M^{jl} + \eta^{jl}M^{ik}
\nn\\
&& -\eta^{il}M^{jk} - \eta^{jk}M^{il} )\, ,
\nn\\[1mm]
[\widehat M^{0i}, M^{jk}]\= -i\,( \eta^{ik}\widehat M^{0j} -
\eta^{ij}\widehat M^{0k} )\, ,
\nn\\[1mm]
[\widehat M^{0i}, \widehat M^{0j}]\= -i\, M^{ij}\, ,
\nn\\[1mm]
[\widehat P^0, \widehat P^0] \=
[\widehat P^0, P^i] \ =\
[ P^i, P^j] \ = \ 0\,,
\nn
\eeq
while the fermionic sector is given by
\beq
\{ \qh_\a , \qbh_\bd\} \= -2\, \left(\s^0_{\a\bd} \widehat P_0
- \s^i_{\a\bd}P_i \right),\nn\\[1mm]
[\widehat M^{0i},\qh_\a] \= -{i\over 2}(\s^{0i})_\a^{\ \b}\, \qh_\b
\,,\nn\\[1mm]
[\widehat M^{0i},\qbh_\ad] \= -{i\over 2}(\bar\s^{0i})_{\ad\bd}\, \qbh^\bd
,\nn\\[1mm]
[M^{ij},\qh_\a] \= -{i\over 2}(\s^{ij})_\a^{\ \b}\, \qh_\b
\,,\nn\\[1mm]
[M^{ij},\qbh_\ad] \= -{i\over 2}(\bar\s^{ij})_{\ad\bd}\, \qbh^\bd,
\nn
\eeq
and
$$
[\widehat P^0,\qh_\a]= [\widehat P^0;\qbh_\ad]  = [P^i;\qh_\a]
 = [P^i,\qbh_\ad] = 0\,.
$$
The {\it thermal super-Poincar\'e algebra} has hence the
same structure as at $T=0$, and contains the same number of supercharges,
once one has appropriately covariantized the generators with respect to
thermal superspace. The thermal time translation operator
$\widehat P^0=-i\widehat\pa^0$ (the thermal covariantization of $P^0$)
can be interpreted as a central charge of the subalgebra one obtains
upon removing  the thermal Lorentz boosts \cite{dl}.

\section{Thermal supersymmetry transformations}\label{ttsstt}

We now compute how the thermal superfield components transform
under the thermal supersymmetry transformations generated by the thermal
supercharges $\qh_\ad$ and $\qbh_\ad$, eqs. \equ{ths}--\equ{thsb}. This means
translating into component language the thermal superfield transformation
$\widehat\delta$ given by
\be
\widehat\delta\widehat\phi(x,\th,\tbh)= i
\left(
\widehat\epsilon^\a\qh_\a +\widehat{\bar\epsilon}_\ad\qbh^\ad
\right)
\widehat\phi(x,\th,\tbh)\,.
\label{tsstrsf}
\ee
The main observation here is that, due to the covariant structure of the
thermal supersymmetry generators, the derivation of the component
transformations is perfectly analogous to the $T=0$ case.
Indeed, the supercharges $\qh_\a$ and $\qbh_\ad$, when acting on
a thermal chiral superfield $\widehat\phi(\yh,\th)$, reduce to
(see \cite{dl} for details)
$$
\ba{rcl}
\qh_\a\widehat\phi(\yh,\th) \= -i \pad{}{\th^\a}\widehat\phi(\yh,\th)
\,,\\[1mm]
\qbh_\ad\widehat\phi(\yh,\th) \= -2(\th\s^\m)_\ad\pad{}{\yh^\m}
\widehat\phi(\yh,\th)\,.
\ea
$$
Inserting this into  \equ{tsstrsf} leads to
$$
\widehat\delta\widehat\phi(\yh,\th)=
\left(
\widehat\epsilon^\a\pad{}{\th^\a}
-2i(\th\s^\m\widehat{\bar\epsilon})\pad{}{\yh^\m}
\right)
\widehat\phi(\yh,\th)\,.
$$
Defining then $\pad{}{\yh^\m}
\varphi(\yh) \equiv \pa_\m \varphi$, for $\varphi=z$ or $\psi$, we get:
\beq
\widehat\delta\widehat\phi(\yh,\th)
\= \widehat\epsilon^\a\left[\sqrt{2}\psi_\a(\yh)
-2\th_\a f(\yh)\right] \nn\\
&&-2i(\th\s^\m\widehat{\bar\epsilon})
\left[\pa_\m z(\yh) +\sqrt{2}\th^\a\pa_\m\psi_\a(\yh)\right]\,.
\nn
\eeq
Comparison with the component expansion of
$\widehat\delta\widehat\phi(\yh,\th)$ immediately leads to:
\be
\ba{rcl}
\widehat\delta z\=
\sqrt{2}\widehat\epsilon^\a \psi_\a \,,
\\[1mm]
\widehat\delta \psi_\a\=
-\sqrt{2}\widehat\epsilon_\a f -i\sqrt{2}
(\s^\m \widehat{\bar\epsilon})_\a (\pa_\m z)\, ,
\\[1mm]
\widehat\delta f
\=-i\sqrt{2} (\s^\m \widehat{\bar\epsilon})^\a (\pa_\m \psi_\a)\,,
\ea
\label{tt2}
\ee
where {\it the unique difference with
the case of zero temperature is the appearance of the thermal
(time-dependent and antiperiodic) spinorial parameter $\widehat\epsilon$,
$\widehat\epsilon(t+i\b)=-\widehat\epsilon(t)$,
in place of the constant spinorial parameter $\epsilon$ of $T=0$
supersymmetry}. The nature of $\widehat\epsilon$  is however deeply
related to finite temperature.
The time-dependence and antiperiodicity of $\widehat\epsilon$ are, in this
thermal superspace formalism, the manifestation of the breaking of global
supersymmetry at finite temperature.

\section{Realizations of thermal supersymmetry -- Wess-Zumino model}
\label{tbbss}

Bosonic and fermionic fields at finite temperature
are characterized by periodic, resp. antiperiodic, boundary
conditions [eqs. \equ{pbc1},\equ{pbc2} and \equ{abc}].
At the level of Green's functions, thermal effects induce
a differentiation between bosons and fermions through the
corresponding KMS conditions \equ{c1}--\equ{c3}, resp. \equ{c2}.
Both the fields' boundary conditions and the KMS
conditions carry information that is of {\it global} character,
in the sense that it relates the values of the field at distant
regions in space-time. This is why the thermal superalgebra,
which is a local structure, is insensitive to such global
conditions and preserves its structure at finite temperature.
In particular, the antiperiodicity conditions on $\th$, $\tbh$,
eq. \equ{antiper}, have no influence on the algebra. It is only the
{\it local} statement that the superspace Grassmann variables
should be allowed to depend on time which makes it
necessary to covariantize the algebra generators.
However, in investigating realizations of the thermal supersymmetry
algebra, we shall be dealing with thermal fields, at the level
of which the thermal  boson/fermion distinctions enter
as global conditions.
Therefore, {\it we expect to see signs of supersymmetry breaking
when realizing the thermal supersymmetry algebra on thermal bosonic
fermionic fields}.  A common way
of introducing the fields' global pe\-rio\-di\-ci\-ty properties is to
develop them thermally {\it \`a la} Matsu\-bara. In the Matsu\-bara
expansion, bosons are expanded in thermal modes as
$$
z(t,\xx)={1\over\sqrt{\b}}\sum_{n} z_n(\xx) \ e^{i\o_n^B t}
\,,\quad \o_n^B= {2n\pi\over \b}
$$
where $\o_n^B$ are the bosonic Matsubara frequencies, while fermions
are developed as
$$
\psi(t,\xx)={1\over\sqrt{\b}}\sum_{n} \psi_n(\xx) \
e^{i\o_n^F t}\,,
\quad \o_n^F= {(2n\! +\!1)\pi\over\b}
$$
with the fermionic Matsubara frequencies $\o_n^F$.
Clearly, these developements contain the information on the periodicity or
antiperiodicity in time, as one may immediately check upon shifting the
time argument by $i\b$.
The Matsubara expansion, after rotation to euclidean time, is a realization
of the imaginary time formalism for finite temperature field theory. It is
an expansion on $S^1\times \real^3$, the circle $S^1$ having length $\beta
= 1/T$. In a supergravity theory, it could be regarded as
a particular Scherk-Schwarz compactification \cite{SchSch} scheme of the
time direction.

Since we have considered only non-interacting scalar and fermionic matter
fields described by chiral and antichiral superfields, the
natural zero-temperature limiting field theory -- to be studied now
at finite temperature -- is the free $T=0$ (off-shell) Wess-Zumino
model
\be
S^{d=4} = \intq \bigl( \LL^{d=4}_{\rm kin} + \LL^{d=4}_{ \rm mass}\bigr)\,,
\label{acto}
\ee
with kinetic and mass lagrangians given by
\beq
\LL^{d=4}_{\rm kin} \=  \half(\pa_\m A)^2 + \half(\pa_\m B)^2 + {i\over 2}
\bar\psi \g^\m\!\pa_\m \psi + \half (F^2\!+\!G^2),\nn\\[1mm]
\LL^{d=4}_{ \rm mass} \=  -M_4(\half \bar\psi \psi + AF + BG)\,.\nn
\eeq
$M_4$ is the mass, $\psi$ a four-component Majorana fermion and
$A$, $B$, $F$, $G$ are real scalar fields which relate to the complex
scalar  $z$ and $f$ through
$$
z(x)={1\over\sqrt{2}}[A(x)+i B(x)]\,,\quad
f(x)={1\over\sqrt{2}}[F(x)+iG(x)]\,.
$$
The supersymmetry transformations at $T=0$ write
\be\ba{rcl}
\d A \= \bar\epsilon \psi\,,
\qquad \d B = i\bar\epsilon\g_5\psi\,,\\[1mm]
\d F \= i\bar\epsilon\g^\m\pa_\m \psi\,,
\qquad \d G =  -\bar\epsilon\g_5
\g^\m\pa_\m \psi\,,\\[1mm]
\d\psi \= -[i\g^\m\pa_\m(A+iB\g_5)+F+iG\g_5]\epsilon\,,
\ea\label{ss2}\ee
and under these, the kinetic and mass
contributions to the action $S^{d=4}$ are separately invariant.
Concretely, omitting in each case a space-time derivative
which integrates to zero,
\be
\label{transf1}
\begin{array}{rcl}
\displaystyle{\d\int d^4x\,\LL^{d=4}_{\rm kin}}
&=&
\displaystyle{\int d^4x\, \overline\psi\gamma^\nu\gamma^\mu[\partial_\mu
(A+iB\gamma_5)]\partial_\nu\epsilon =0\,,}
\\[2mm]
\displaystyle{\d\int d^4x\,\LL^{d=4}_{ \rm mass}} &=&
\displaystyle{-iM_4 \int d^4x\, \overline\psi\gamma^\mu(A+iB\gamma_5)
\partial_\mu\epsilon=0 }\,,
\end{array}
\ee
which of course vanish at zero temperature where $\epsilon$ is a
constant spinor.

Performing the thermal expansion of \equ{acto}
(see \cite{dl} for details), we get
the $d=3$  euclidean expression
\beq
S^{d=3}
\=
\displaystyle{\!\intt\!\!\! \sum_{n}}\!
\Biggl\{
\displaystyle\half \pa^i A^*_{n} \pa_i A_n
+\displaystyle\half \pa^i B^*_{n} \pa_i B_n\nn\\
&+&\displaystyle\half(M^B_{3,n})^2(A^*_{n} A_n\!+\!B^*_{n} B_n)
+\displaystyle\half\biggl[\l_n^\dagger (\s^i\pa_i\! -\!\o_n^F)\l_n\nn\\
&+&
 M_4  \l_{-n-1}^\dagger i\s^2\l^*_n\biggr] +{\rm h.c.}\Biggr\}\,,
\label{mumu}
\eeq
where the thermal mass of the $n$-th $d=3$ bosonic mode $M^B_{3,n}$ obeys
$$
(M^B_{3,n})^2  = M_4^2 + (\o_n^B)^2,
\qquad  (\o_n^B)^2={4\pi^2 n^2\over\b^2}\,.
$$
For fermions, the mass matrix in \equ{mumu} can be written:
$$
{1\over2}\sum_n
\left( \lambda^\dagger_n \quad \lambda_{-n-1}\right)
\left( \begin{array}{cc} \omega_n^F & M_4 \\
M_4 & -\omega_n^F  \end{array}\right)
\left( \begin{array}{c} \lambda_n \\ \lambda^\dagger_{-n-1}
\end{array} \right)
+ {\rm h.c.}
$$
and posesses two opposite eigenvalues $\pm M_{3,n}^F$  verifying
$$
(M^F_{3,n})^2  = M_4^2 + (\o_n^F)^2\,,\qquad  (\o_n^F)^2={\pi^2
(2n+1)^2 \over
\b^2}\,.\label{3fm}
$$
{}From these mass relations, it is clear
that {\it thermal effects lift the mass degeneracy characteristic of $T=0$
supersymmetry}. The lifting of the mass degeneracy by temperature effects
 is a clear signature of thermal supersymmetry
breaking at the level of thermal fields.

We also expect to see thermal supersymmetry breaking when trying to realize
the thermal supersymmetry algebra on systems of thermal fields.	In order to
investigate this, we first need to pin down the thermal supersymmetry
transformations of the component fields, as we have expressed our Wess-Zumino
model in components.
In Section \ref{ttsstt}, we have shown that component transformations under
thermal supersymmetry have the same form as at $T=0$, but with the space-time
constant supersymmetry parameter $\epsilon$ replaced by the
thermal, time-dependent and antiperiodic quantity $\widehat\epsilon$.
This allows us  to identify immediately the thermal
version of the transformations \equ{ss2} as
\beq
\widehat\d A
\=
 \widehat{\bar\epsilon} \psi\,,
\qquad
\widehat\d B
=
  i\widehat{\bar\epsilon}\g_5\psi
\,,\nn
\\[1mm]
\widehat\d F
\=
 i\widehat{\bar\epsilon}\g^\m (\pa_\m \psi)\,,
\qquad
\widehat\d G
=
  -\widehat{\bar\epsilon}\g_5 \g^\m(\pa_\m \psi)
\,,\label{ss1}
\\[1mm]
\widehat\d\psi
\=
 -[i\g^\m(\pa_\m(A+iB\g_5))+F+iG\g_5]\widehat\epsilon\,.
\nn
\eeq
These expressions can be translated into transformations of the
three-dimensional Matsubara modes, as is shown in \cite{dl}. When doing so,
one must take into account the fact that, the supersymmetry parameters
being now time-dependent, they must be developed thermally (with constant
modes). One therefore obtains, at the level of the $d=3$ thermal modes,
thermal supersymmetry transformations with parameters which carry an index
of the modes, and as a consequence mix the modes of bosons and fermions
\cite{dl}. This should be contrasted with what one has at zero temperature,
in which case the supersymmetry parameters are space-time constants that
one does  not develop thermally when dimensionally reducing to $d=3$.

The thermal action has -- in contrast to \equ{transf1} -- the following
non trivial variation under thermal supersymmetry (\ref{ss1}) [a rotation
to imaginary (euclidean) time is understood]:
\be
\label{transf2}
\begin{array}{rcl}
\displaystyle{\widehat\d\int d^4x\,\LL^{d=4}_{\rm kin}}
&=&
\displaystyle{\int d^4x\, \psi^\dagger\gamma^\mu\partial_\mu
(A+iB\gamma_5)\partial_0\widehat\epsilon\,,}
\\[2mm]
\displaystyle{\widehat\d\int d^4x\,\LL^{d=4}_{ \rm mass}}
&=&
\displaystyle{-iM_4 \int d^4x\, \psi^\dagger(A+iB\gamma_5)
\partial_0\widehat\epsilon }\,,
\end{array}
\ee
where $\partial_0\widehat\epsilon$ does  {\it not} vanish as
$\widehat\epsilon$ depends on time.
Clearly, neither the kinetic action nor the mass action are
invariant under the thermal supersymmetry transformations.
Upon inserting the Matsubara mode expansions, these variations
can be translated into {\it temperature-dependent}, $d=3$
expressions \cite{dl}.
The variation of the total action is then seen to be
proportional to $\o_n^F \sim T$, as a consequence of thermal
supersymmetry breaking.
In the $T\rightarrow 0$ limit, one expects supersymmetry
to be thermally unbroken. The variations $\d S^{d=4}_{\rm kin}$
and $\d S^{d=4}_{\rm mass}$ indeed vanish separately in
that limit.

\section{Conclusions}\label{concl}

Immersing a physical system in a heat bath results in the fields
acquiring different properties according to their statistics.
{\it E.g.}, finite-temperature bosonic fields obey periodic boundary
conditions, while fermionic fields satisfy antiperiodic b.c.'s.
Such a distinction can be seen also at the level of the Green's functions.
Depending on their statistics, thermal propagators obey either a
bosonic KMS condition, or a fermionic one. Therefore, thermal
effects induce a clear and mandatory distinction between bosons and
fermions. As a consequence, finite temperature environments are
incompatible with $T=0$ supersymmetry : the supersymmetry transformations
are indeed unable to take into account the distinct thermal properties
that go along with different statistics.

The approach we propose (see also \cite{dl}) allows to formulate
(broken) supersymmetry in a thermal environment in such a way as to respect
the different thermal behaviours of bosons and fermions. The
parameters of supersymmetry transformations being antiperiodic
at finite temperature, as advocated in \cite{ggs}, thermal supersymmetry
transforms periodic bosons into antiperiodic fermions, and {\it vice-versa}.
At the level of superfields, our approach makes it possible to
reconcile the component fields' distinct boundary conditions
within a {\it superfield} boundary condition. For chiral superfields,
this b.c. is of bosonic type and is formulated in thermal superspace.
The latter is spanned by usual space-time and by time-dependent and
antiperiodic Grassmann coordinates, with a period given by the inverse
temperature. The superspace Grassmann coordinates are thus subject
to a temperature-dependent constraint similar to that obeyed by
the supersymmetry parameters.

Thermal superfield propagators are shown to obey a KMS
condition formulated directly at the level of superfields.
Its proof makes an essential use of the antiperiodicity of the
Grassmann coordinates. This antiperiodicity is crucial as well
in proving that the superfield KMS condition implies the correct,
bosonic or fermionic, KMS condition for the superfield components.

In this sense, the thermal superspace approach allows to reconcile
the notions of finite temperature physics and of supersymmetry,
yielding {\it a formalism for bosons and fermions in interaction with
a heat bath in which thermal supersymmetry breaking is encoded}.
Such a formalism is particularly welcome for, \eg cosmology.
Thermal superspace is therefore the  correct superspace for finite
temperature situations. It is shown to provide a natural representation
for the thermal supersymmetry algebra, which is obtained upon covariantizing
thermally the supercharges as well as the Lorentz and translation generators.
The thermal supersymmetry algebra has the same structure as at $T=0$ and
the same number of supersymmetries, while thermal supersymmetry breaking is
encoded
in the temperature-dependent conditions we impose on superspace.

It is only when trying to realize the thermal supersymmetry algebra on systems
of
thermal fields that one encounters explicit thermal supersymmetry breaking.
Indeed, the conditions which distinguish bosons from fermions at finite
temperature
-- the fields' b.c.'s or the KMS conditions -- carry information that is of
{\it global} nature in space-time\footnote{Previous studies of thermal
supersymmetry
breaking, \eg \cite{dk,ggs,fujikawa,boyan,gs,bo} have  been conducted at the
level
of thermal fields/states with this global periodicity/antiperiodicity
distinction.
Note that the discussion of the thermal breaking of supersymmetry is closely
related
to that of the thermal breaking of Lorentz invariance \cite{aoyama,ojima}, for
which
a Lorentz covariant treatment has been given \cite{weldon}.}.
The supersymmetry algebra, being a {\it local} structure, is insensitive
to this global information. It only needs to be covariantized with respect
to the  {\it local} statement that the superspace Grassmann parameters depend
on imaginary time.
At the level of thermal actions, we encounter thermal supersymmetry breaking
in two ways. First, upon developping thermally the $T=0$ Wess-Zumino model, we
observe that the mass degeneracy is lifted. Second, we compute the variation of
the thermal action under thermal supersymmetry and observe that the action is
non-invariant. The variation is given, in terms of thermal modes, by a
temperature-dependent expression which vanishes in the $T\rightarrow 0$ limit
where one expects supersymmetry to be thermally unbroken.

\vspace{3mm}

\noindent{\bf Acknowledgements}: The author is indebted to the organisers
of the 1998 Miramare Summer Institute and to SISSA, Trieste, for hospitality
during the redaction of these proceedings.



\begin{thebibliography}{99}


\bibitem{dl} J.-P. Derendinger and C. Lucchesi,
hep-ph/9807403.

\bibitem{dk} A. Das and M. Kaku,
{\it Phys. Rev.} {\bf D18} (1978) 4540.

\bibitem{ggs} L. Girardello, M.T. Grisaru and P. Salomonson,
{\it Nucl. Phys.} {\bf B178} (1981) 331.

\bibitem{fujikawa} K. Fujikawa,
{\it Z. Phys.} {\bf C15} (1982) 275.

\bibitem{vh} L. van Hove,
{\it Nucl. Phys.} {\bf B207} (1982) 15;
T.E. Clark and S.T. Love,
{\it Nucl. Phys.} {\bf B217} (1983) 349;
D.A. Dicus and X.R. Tata,
{\it Nucl. Phys.} {\bf B239} (1984) 237;
A. Cabo, M. Chaichian and A.E. Shabad,
{\it Phys. Lett.} {\bf B294} (1992) 217.

\bibitem{fuchs} J. Fuchs,
{\it Nucl. Phys.} {\bf B246} (1984) 279.

\bibitem{boyan} D. Boyanovski,
{\it Phys. Rev.} {\bf D29} (1984) 743;
H. Aoyama and D. Boyanovski,
{\it Phys. Rev.} {\bf D30} (1984) 1356;
 D. Boyanovski,
{\it Physica} {\bf 15D} (1985) 152.

\bibitem{susysound} R. Godmundsdottir 
and P. Salomonson,  {\it Phys. Lett.}
{\bf B182} (1986) 174;  V.V. Lebedev 
and A.V. Smilga, {\it Nucl. Phys.} {\bf
B318} (1989) 669.

\bibitem{unbroken} S.P. Chia,
{\it Phys. Rev.} {\bf D33} (1986) 2481;  {\it ibid.} 2485;
S. Kumar,
{\it J. Phys. A: Math. Gen.} {\bf 23} (1990) L127;

\bibitem{gs}
R. Godmundsdottir and P. Salomonson,
{\it Nucl. Phys.} {\bf B285} [FS19] (1987) 1.

\bibitem{d} A. Das,
{\it Physica} {\bf A158} (1989) 1.

\bibitem{hight} W.-H. Kye, S.K. Kuang and J.K. Kim,
{\it Phys. Rev.} {\bf D46} (1992) 1835;
H.-S. Song, G.-N. Xu and I. An,
{\it J. Phys. A: Math. Gen.} {\bf 26} (1993) 4463

\bibitem{lr} R.G. Leigh and R. Rattazzi,
{\it Phys. Lett.} {\bf B352} (1995) 20.

\bibitem{bo} D. Buchholtz and I. Ojima,
hep-th/9701005.

\bibitem{kms} R. Kubo,
{\it J. Phys. Soc. Japan} {\bf 12} (1957) 570;
P.C. Martin and J. Schwinger,
{\it Phys. Rev.} {\bf 115} (1959) 1342.


\bibitem{SchSch} J. Scherk and J.H. Schwarz,
{\it Phys. Lett.} {\bf B82} (1979) 60;
E. Cremmer and J. Scherk,
{\it Nucl. Phys.} {\bf B103} (1976) 399.

\bibitem{aoyama} H. Aoyama,
{\it Phys. Lett.} {\bf B171} (1986) 420.

\bibitem{ojima} I. Ojima,
{\it  Lett. Math. Phys.} {\bf 11} (1986) 73.

\bibitem{weldon} H.A. Weldon,
{\it Phys. Rev.} {\bf D26} (1982) 1394.

\end{thebibliography}
\end{document}